\documentclass{article}
\usepackage{arxiv}
\usepackage[utf8]{inputenc}
\usepackage[T1]{fontenc}
\usepackage{url}
\PassOptionsToPackage{hyphens}{url}
\usepackage[colorlinks=true,linkcolor=blue,urlcolor=blue,citecolor=blue,breaklinks=true]{hyperref}

\usepackage{booktabs}
\usepackage{amsfonts}
\usepackage{nicefrac}
\usepackage{microtype}
\usepackage{lipsum}
\usepackage{graphicx}
\usepackage{algorithm}
\usepackage{algpseudocode}
\usepackage{pgfplots}
\usepackage{subcaption}
\usepackage{enumitem}
\usepackage{pgfplotstable}
\pgfplotsset{compat=1.18}
\graphicspath{ {./images/} }

\title{Cross-Technology Generalization in Synthesized Speech Detection: Evaluating AST Models with Modern Voice Generators}

\author{
Andrew Ustinov \\
Department of Machine Learning and Data Analytics\\
ITMO University\\ 
AI Innovation Hub\\
ANTEI Limited\\
\texttt{entro1122@gmail.com} \\
\And
Matey Yordanov \\
Department of Integrative Systems and Design\\
Hong Kong University of Science and Technology\\ 
AI Innovation Hub \\
ANTEI Limited \\
\texttt{msyordanov@connect.ust.hk} \\
\And
Andrei Kuchma\\
Institute of Applied Computer Science \\
ITMO University\\
AI Innovation Hub\\
ANTEI Limited \\
\texttt{akuchma@anteihk.com} \\
\And
Mikhail Bychkov\\
Department of Risk Management \& Business Intelligence \\
Hong Kong University of Science and Technology\\
AI Innovation Hub\\
ANTEI Limited \\
\texttt{mbychkov@anteihk.com} \\
}

\begin{document}
\maketitle

\begin{center}
\textbf{Abstract}
\end{center}

This paper evaluates the Audio Spectrogram Transformer (AST) architecture for synthesized speech detection, with focus on generalization across modern voice generation technologies. Using differentiated augmentation strategies, the model achieves 0.91\% EER overall when tested against ElevenLabs, NotebookLM, and Minimax AI voice generators. Notably, after training with only 102 samples from a single technology, the model demonstrates strong cross-technology generalization, achieving 3.3\% EER on completely unseen voice generators. This work establishes benchmarks for rapid adaptation to emerging synthesis technologies and provides evidence that transformer-based architectures can identify common artifacts across different neural voice synthesis methods, contributing to more robust speech verification systems.


\section{Introduction}
Americans lost nearly \$9 billion to deepfake fraud last year alone – an increase of over 150\% in just two years, according to the Federal Trade Commission\cite{ftc2024}. This statistic reflects the growing threat associated with the malicious use of modern generative models. Market analysis indicates that generative AI could enable fraud losses to reach \$40 billion in the United States by 2027, up from \$12.3 billion in 2023, representing a compound annual growth rate of 32\%\cite{realitydefender2024}. This projection significantly understates the true scale, as many incidents remain unreported due to reputational concerns or detection failures. The global losses to all forms of scams have surpassed \$1.026 trillion, equivalent to 1.05\% of global GDP\cite{newswire2024}.

Deepfake fraud has inflicted severe financial damage across industrial sectors. Research shows that 92\% of surveyed businesses have experienced losses up to \$450,000 from deepfake fraud, with 10\% reporting losses exceeding \$1 million\cite{regula2024}. The financial services sector bears a disproportionate burden, with average losses exceeding \$603,000 per incident\cite{regula2024}. Further breakdown reveals that fintech organizations experience average losses of \$637,000, while traditional banking institutions report \$570,000 in average damages per attack\cite{regula2024}.

The economic impact shows significant geographic variation. Mexico reported the highest average losses at \$627,000 per incident, followed by Singapore (\$577,000), the United States (\$438,000), Germany (\$394,000), and the UAE (\$379,000)\cite{regula2024}. In Australia alone, nearly 240,000 citizens reported being victims of voice cloning scams in 2022, resulting in financial losses of A\$568 million\cite{conversation2024}. This global distribution of attacks underscores the universal nature of the threat.

The proliferation of audio deepfakes extends beyond individual consumers to systematically target organizations. Recent survey data reveals that 49\% of businesses faced audio and video deepfake attacks in the past 12 months, with audio deepfake incidents increasing by 12\% compared to previous reporting periods\cite{biometric2024}. Sector-specific analysis demonstrates that audio deepfakes are particularly prevalent in financial services (51\%) and cryptocurrency operations (55\%), highlighting the financial motivation behind many of these attacks\cite{biometric2024}.

The accessibility of this technology presents a significant concern. Modern software tools can recreate a person's voice after analyzing a sample as short as 3 seconds\cite{regula2024}, making this technology particularly dangerous in an era of ubiquitous audio and video content on social media. Common voice cloning solutions require only 10 minutes of learning to reproduce convincing voice replicas\cite{ftc2024}.

Detection capabilities remain woefully inadequate against this growing threat. Human listeners demonstrate only 53.7\% accuracy in identifying synthetic audio, a figure expected to decrease as AI technology continues to advance\cite{pindrop2024}. Evaluations of 14 publicly available detection tools concluded that current audio deepfake detectors cannot be relied upon for practical use\cite{northwestern2024}. This detection challenge is compounded by the increasing sophistication of attack methodologies, with large-scale operations like the recently discovered "FM Scam" spoofing over 500,000 devices and generating up to 100 million fraudulent ad requests monthly\cite{doubleverify2024}.

The financial damage from AI-generated voice fraud significantly exceeds losses from traditional phone scams. While conventional fraud calls in the United States result in average losses of \$539 per victim, deepfake-powered scams frequently lead to damages exceeding \$6,000 per incident\cite{businesswire2025}. This sixfold increase in financial impact demonstrates the heightened effectiveness of AI-powered voice manipulation in circumventing existing security measures. Furthermore, 51.6\% of C-suite executives anticipate an increase in both the frequency and magnitude of deepfake attacks targeting their organizations' financial systems in the coming year\cite{realitydefender2024}.

While detection of fake images has received significant attention from the research community, the audio domain remains insufficiently studied. This disparity is partially explained by economic factors—there exist considerable financial incentives for creating deepfakes but limited economic motivation for developing detection solutions\cite{berkeley2024}. 

The main objective of this research is to develop effective methods for classifying and detecting synthesized speech, which is critical for protecting users from financial damage associated with such attacks. Our work aims to address this gap by creating and analyzing a dataset consisting of samples from several state-of-the-art neural network architectures for speech synthesis, contributing to the development of more robust protective measures against this emerging threat.

\section{Research Objectives}

The primary objective of this research is to comprehensively evaluate the performance and generalization capabilities of Audio Spectrogram Transformer (AST) architecture \cite{gong2021ast} in the domain of synthesized speech detection. This investigation encompasses several specific aims:

\begin{enumerate}
    \item To assess the model's adaptability using a focused custom dataset comprising samples from the three most sophisticated contemporary voice synthesis technologies: ElevenLabs, NotebookLM, and Minimax AI generators, addressing the urgent need for evaluation against these advanced synthetic speech technologies.
    
    \item To develop differentiated augmentation strategies for bonafide and synthetic speech samples that acknowledge their inherent acoustic disparities, thus enhancing the model's ability to identify persistent artifacts within synthetic speech rather than dataset-specific characteristics.
    
    \item To evaluate the rapid adaptation capabilities of AST models by measuring how effectively they can identify patterns from modern voice generators after minimal exposure, using only 102 samples from a single synthetic voice technology during training, compared to thousands of data entries required for training by other deepfake detection algorithms.
    
    \item To analyze cross-technology generalization capabilities by training exclusively on one synthetic voice technology (ElevenLabs) and testing on others (NotebookLM and Minimax AI), determining whether common artifacts exist across different neural voice synthesis methods.
    
    \item To establish benchmarks for efficient adaptation to emerging voice synthesis technologies, with particular emphasis on minimizing the quantity of samples required to achieve effective detection performance against previously unseen generation techniques.
\end{enumerate}

This research addresses a significant gap in the literature regarding how effectively detection systems can generalize to novel voice synthesis technologies. Few public studies have examined model performance against the latest generators such as ElevenLabs, NotebookLM, and Minimax AI. The evaluation of AST architecture demonstrates promising results with minimal training examples, contributing valuable insights for developing robust verification systems that can rapidly adapt to emerging synthetic speech technologies. These findings have practical implications for real-world deployment scenarios where new generators frequently emerge and comprehensive training data may be limited.

\section{Architecture}
\subsection{Audio Spectrogram Transformer Architecture}

This research utilizes the Audio Spectrogram Transformer (AST) architecture, which represents the first fully attention-based model for audio classification that does not employ convolutional layers \cite{gong2021ast}. AST applies a transformer directly to audio spectrograms, allowing it to capture global context even in the lowest layers.

\subsubsection{Operating Principle}
The incoming audio signal of $t$ seconds is converted into a sequence of 128-dimensional log Mel filterbank features, computed using a 25 ms Hamming window with a 10 ms step \cite{gong2021ast, vaswani2017attention}. This results in a spectrogram of size $128 \times 100t$, which serves as the input to AST.

The spectrogram is divided into a sequence of $N$ patches of size $16 \times 16$ with an overlap of 6 pixels in both dimensions (time and frequency), where $N = 12\lceil(100t - 16)/10\rceil$ is the number of patches \cite{gong2021ast, yuan2021tokens}. Each $16 \times 16$ patch is transformed into a one-dimensional embedding of dimension 768 using linear projection. This operation is analogous to the Vision Transformer (ViT) approach used for image processing \cite{dosovitskiy2021image}.

Since the transformer architecture does not account for the order of input data, and the sequence of patches does not correspond to temporal order, a learnable positional embedding (also of dimension 768) is added to each patch embedding, allowing the model to consider the spatial structure of the 2D spectrogram \cite{vaswani2017attention, devlin2019bert}.

A special classification token [CLS] is added to the beginning of the sequence, similar to the BERT architecture \cite{devlin2019bert}. The resulting sequence is then fed into the transformer encoder. AST uses only the transformer encoder since the model is designed for classification tasks.

\begin{figure}[ht]
\centering
\begin{minipage}{0.5\textwidth}
  \centering
  \includegraphics[width=\linewidth]{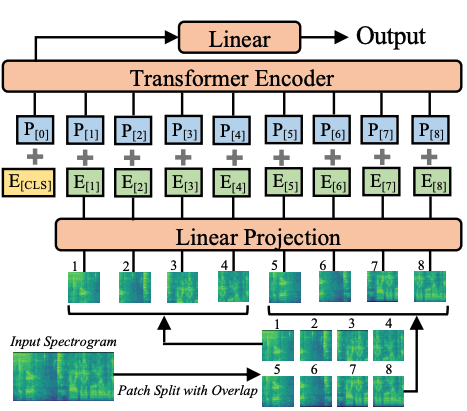}
  \label{fig:first}
\end{minipage}\hfill
\end{figure}

\subsubsection{Hyperparameters and Architectural Features}
The transformer encoder used in AST has the following characteristics \cite{gong2021ast}:
\begin{itemize}
    \item Embedding dimension: 768
    \item Number of layers: 12
    \item Number of attention heads: 12
    \item Patch size: $16 \times 16$
    \item Patch overlap: 6 pixels in both dimensions
    \item Activation function in the output layer: sigmoid (for multi-label classification)
\end{itemize}

These parameters correspond to the standard Transformer encoder architecture as described in the original Transformer paper \cite{vaswani2017attention}. The AST architecture intentionally utilizes the original Transformer encoder without modifications to maintain simplicity and reproducibility \cite{gong2021ast}.

\subsubsection{AST Performance Characteristics}
One notable characteristic of AST is its superior performance across various audio classification benchmarks. When evaluated on AudioSet \cite{gemmeke2017audio}, ESC-50 \cite{piczak2015esc}, and Speech Commands V2 \cite{warden2018speech}, AST achieves state-of-the-art results with mAP of 0.485 on AudioSet, 95.6\% accuracy on ESC-50, and 98.1\% accuracy on Speech Commands V2 \cite{gong2021ast}.

The AST architecture offers several advantages over CNN-attention hybrid models \cite{kong2020panns, gong2021psla}. First, AST naturally supports variable-length inputs and can be applied to different tasks without any architectural changes. The models used for all aforementioned tasks have the same architecture while the input lengths vary from 1 second (Speech Commands) to 10 seconds (AudioSet) \cite{gong2021ast}. In contrast, CNN-based models typically require architecture tuning for different tasks.

Additionally, AST features a simpler architecture with fewer parameters compared to state-of-the-art CNN-attention hybrid models. This simplicity contributes to faster convergence during training, with AST requiring significantly fewer training epochs than hybrid models \cite{gong2021ast}.

The multi-head self-attention mechanism in AST allows the model to focus on different parts of the input spectrogram simultaneously, which is particularly beneficial for audio event detection where relevant information may be scattered across the time-frequency representation \cite{vaswani2017attention, gulati2020conformer}. This ability to capture long-range dependencies in audio signals makes AST particularly well-suited for complex audio classification tasks.

\section{Approach}

\subsection{Methodology and Implementation Details}

This section presents the experimental approach to evaluating the AST model for deepfake speech detection across standard benchmarks and contemporary voice synthesis technologies.

\subsubsection{Data Augmentation Strategy}

A key innovation in the approach is the implementation of distinct augmentation strategies for bonafide and synthetic speech samples. This differentiation acknowledges the inherent acoustic disparities between human and synthetically generated speech, allowing for more targeted model training.

For synthetic speech samples, a comprehensive augmentation pipeline was applied to simulate various acoustic environments and transmission channels that might be encountered in real-world fraud scenarios:

\begin{itemize}
    \item \textbf{Room Simulation}: Virtual acoustic environments with dimensions ranging from 2.0 to 12.0 meters in width and length, and 2.0 to 5.0 meters in height were modeled. Absorption coefficients varied between 0.05 and 0.5, simulating different room materials. This augmentation was applied with probability $p=0.1$.
    
    \item \textbf{Temporal Masking}: Random segments comprising 5\% to 15\% of the audio duration were masked, simulating packet loss or brief interruptions in transmission ($p=0.3$).
    
    \item \textbf{Pitch Shifting}: Subtle pitch modifications within $\pm$1 semitone range were introduced to simulate natural variations in voice production or minor processing artifacts ($p=0.3$).
    
    \item \textbf{Parametric Equalization}: A seven-band parametric equalizer modified frequency components with gain adjustments between -12dB and +12dB, simulating different microphone characteristics and transmission channels ($p=0.3$).
    
    \item \textbf{Gaussian Noise Addition}: Background noise with amplitude ranging from 0.0005 to 0.01 was added, simulating various environmental conditions ($p=0.5$).
    
    \item \textbf{Signal-to-Noise Ratio Variation}: Gaussian noise was added while maintaining SNR between 10dB and 40dB ($p=0.5$).
    
    \item \textbf{Bandpass Filtering}: Audio was filtered to pass frequencies between a center frequency range of 200Hz to 5000Hz, with bandwidth fraction between 0.4 and 1.8 ($p=0.1$).
    
    \item \textbf{Lowpass Filtering}: Frequencies above a cutoff point between 2000Hz and 7500Hz were attenuated, simulating telephone or compressed audio transmission ($p=0.1$).
    
    \item \textbf{Gain Adjustment}: Overall volume was modified between -10dB and +10dB ($p=0.5$).
    
    \item \textbf{Dynamic Gain Transition}: Temporal gain changes from -10dB to +3dB over durations of 0.3 to 0.7 seconds were introduced, simulating automatic gain control in recording devices ($p=0.4$).
    
    \item \textbf{MP3 Compression}: Audio was compressed using MP3 encoding with bitrates varying from 32kbps to 192kbps, simulating distribution conditions for deepfake content ($p=0.4$).
\end{itemize}

For bonafide speech samples, a more conservative augmentation strategy was implemented to preserve natural speech characteristics while still accounting for real-world recording variations:

\begin{itemize}
    \item \textbf{Moderate Room Simulation}: Smaller room dimensions (2.0-8.0 meters) with more moderate absorption coefficients (0.1-0.3) were used with lower probability ($p=0.3$).
    
    \item \textbf{Subtle Noise Addition}: Lower amplitude Gaussian noise (0.0005-0.005) was added with reduced probability ($p=0.3$).
\end{itemize}

\begin{figure}[ht]
\centering
\begin{minipage}{1\textwidth}
  \centering
  \includegraphics[width=\linewidth]{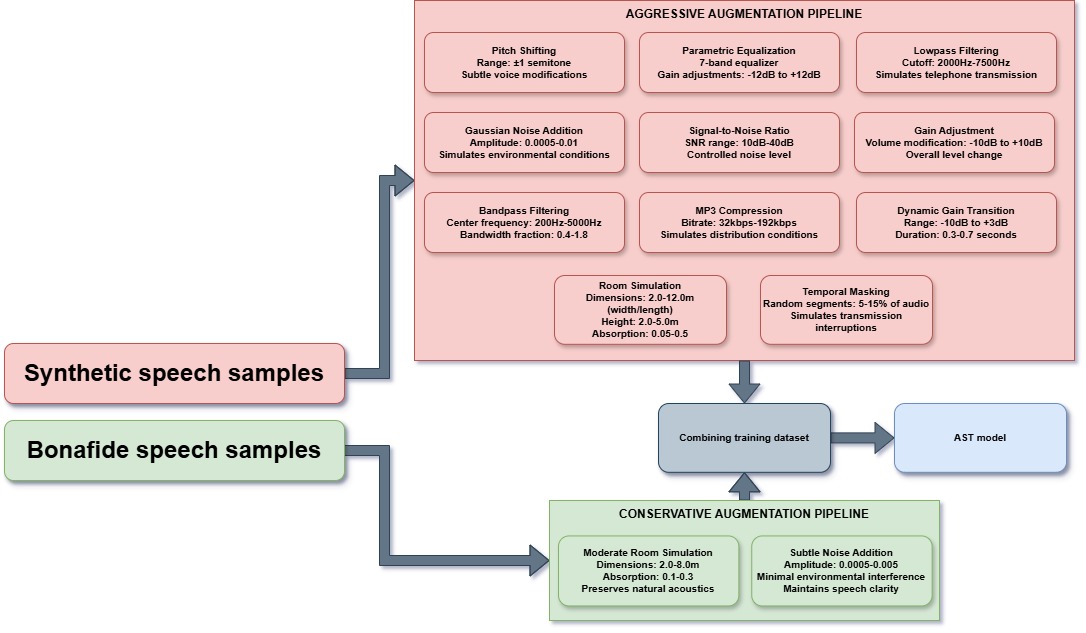}
  \label{fig:first}
\end{minipage}\hfill
\end{figure}

This dual-augmentation approach balances two competing objectives: 1) ensuring the model learns to distinguish synthetic speech artifacts rather than dataset-specific characteristics, and 2) preserving the natural variability present in authentic human speech. By applying more aggressive augmentations to synthetic samples, the hypothesis is that the model will focus on persistent artifacts that remain identifiable despite signal degradation.

\subsubsection{Training Configuration}

The AST model was trained with the following hyperparameters:

\begin{itemize}
    \item \textbf{Mixed Precision}: FP16 computation was utilized to accelerate training while maintaining numerical stability.
    
    \item \textbf{Training Duration}: Maximum 2500 optimization steps with early stopping based on validation performance.
    
    \item \textbf{Batch Size}: 16 samples per device for training and 32 samples per device for evaluation.
    
    \item \textbf{Weight Decay}: 0.02 to prevent overfitting and improve generalization.
        
    \item \textbf{Validation Strategy}: Model performance evaluated every 1000 steps on the validation set.
    
    \item \textbf{Model Selection}: Best model selected based on Equal Error Rate (EER) on the validation set.
\end{itemize}

All audio samples were resampled to 16kHz prior to feature extraction to ensure consistent processing across datasets with varying native sampling rates. The model output was configured as a binary classification with a two-dimensional logit vector, where the first dimension represents synthetic speech probability and the second dimension represents bonafide speech probability.

\section{Experimental Results}

\subsection{Performance on Standard Benchmarks}

This approach demonstrates significant improvements over existing methods on the ASVspoof2019 LA dataset. Table \ref{tab:asvspoof_comparison} presents a comparative analysis of the AST model against previous state-of-the-art approaches.

\begin{table}[h]
\centering
\caption{Equal Error Rate (EER) Comparison on ASVspoof2019 LA Dataset}
\label{tab:asvspoof_comparison}
\begin{tabular}{lc}
\toprule
Model & ASVspoof2019 LA (EER \%) \\
\midrule
LCNN & 6.35 \\
MesoNet & 7.42 \\
MesoInception & 10.02 \\
ResNet18 & 6.55 \\
RawPC & 3.09 \\
RawNet2 & 3.15 \\
RawGAT-ST & 1.23 \\
\textbf{AST with Mixed Training} & \textbf{2.75} \\
\bottomrule
\end{tabular}
\end{table}

By incorporating a portion of the custom dataset with the ASVspoof2019 training data, the enhanced AST model achieved an EER of 2.75\%, representing a significant improvement. This enhancement demonstrates the value of incorporating diverse synthetic speech samples from contemporary voice generation technologies into the training process.

\subsection{Custom Dataset Composition and Evaluation}

To evaluate the generalization capabilities of the model against the most recent voice synthesis technologies, a custom dataset was constructed comprising samples from multiple sources:

\begin{enumerate}
    \item \textbf{Bonafide Speech}: 5000 real human voice samples from the Mozilla Common Voice dataset
    \item \textbf{Synthetic Speech}:
    \begin{itemize}
        \item 1029 samples from ElevenLabs voice synthesis
        \item 257 samples from NotebookLM podcast voice synthesis
        \item 59 samples from Minimax AI voice synthesis technology
    \end{itemize}
\end{enumerate}

The dataset was split to evaluate the model's ability to generalize from limited exposure to new synthetic speech technologies:

\begin{itemize}[label={•}, resume, leftmargin=*]
    \item \textbf{Training Set (9.7\% of total data)}:
    \begin{itemize}
        \item 500 real human voice samples
        \item 102 ElevenLabs synthetic samples
    \end{itemize}
    
    \item \textbf{Testing Set (90.3\% of total data)}:
    \begin{itemize}
        \item 4500 real human voice samples
        \item 927 ElevenLabs synthetic samples
        \item 257 NotebookLM synthetic samples
        \item 59 Minimax AI synthetic samples
    \end{itemize}
\end{itemize}

This configuration enables assessment of both within-distribution performance (ElevenLabs) and out-of-distribution generalization (NotebookLM and Minimax AI).

\begin{figure}[ht]
\centering
\begin{minipage}{0.5\textwidth}
  \centering
  \includegraphics[width=\linewidth]{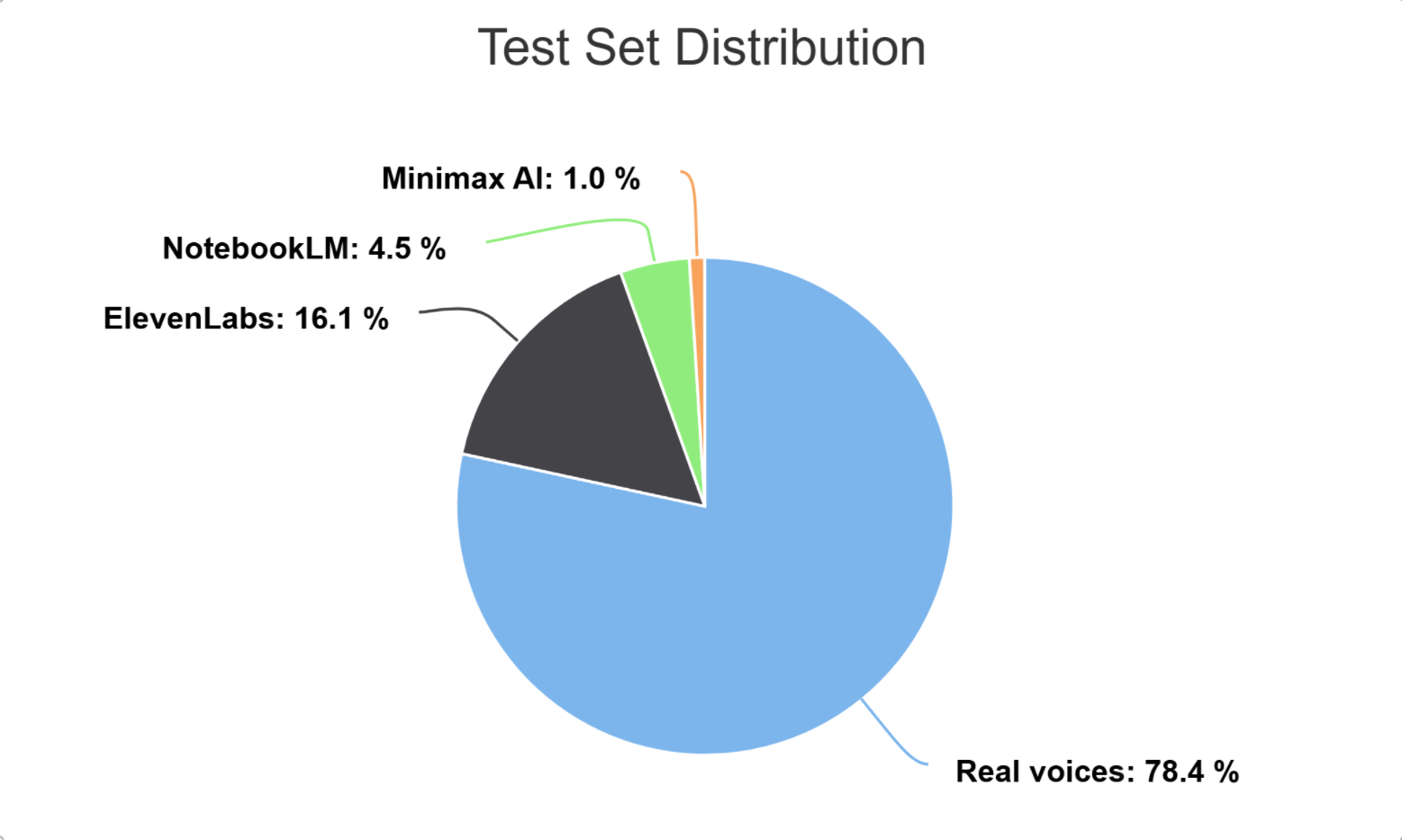}
  \label{fig:first}
\end{minipage}\hfill
\begin{minipage}{0.5\textwidth}
  \centering
  \includegraphics[width=\linewidth]{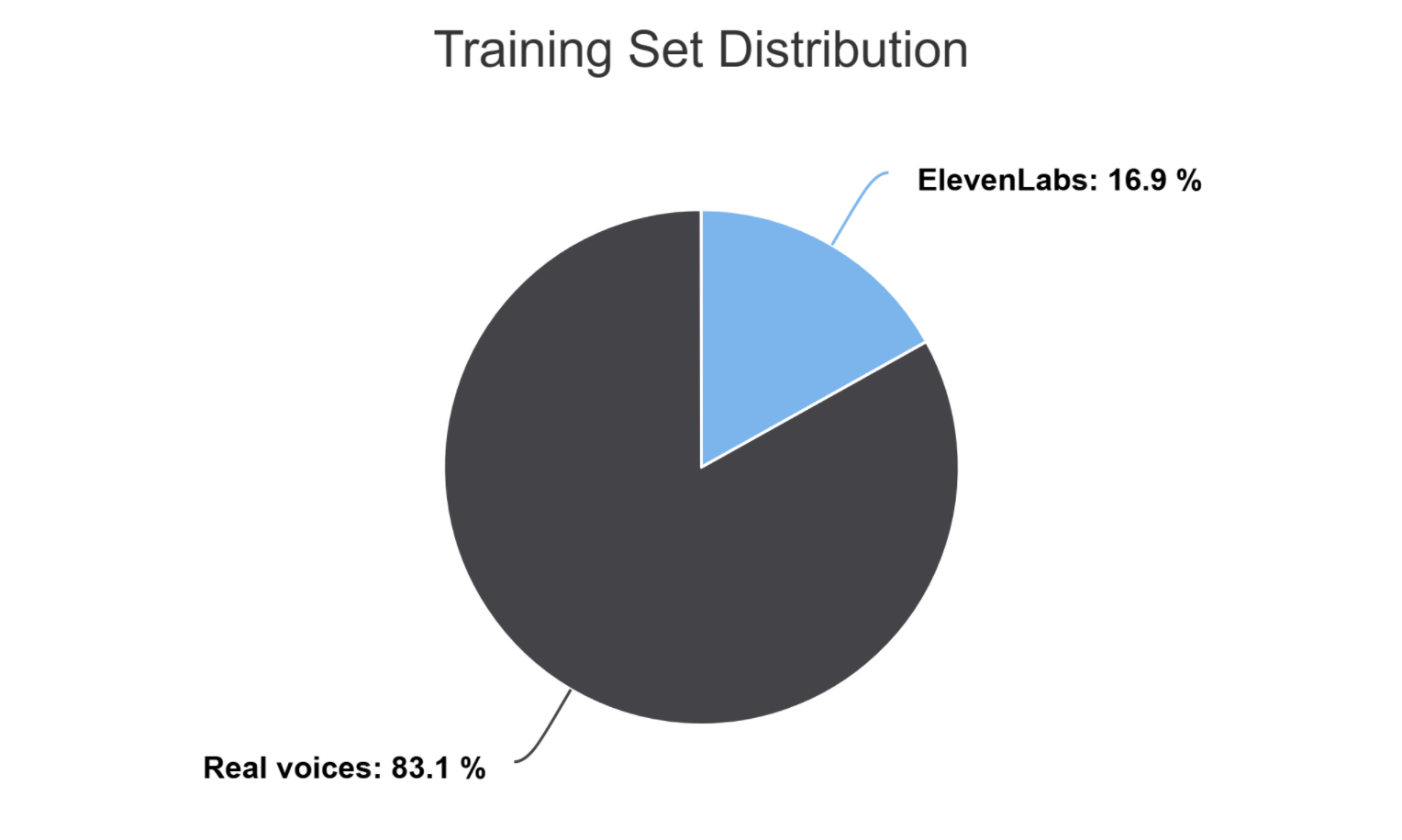}
  \label{fig:second}
\end{minipage}
\end{figure}

\subsection{Cross-Technology Generalization Results}

The AST model, which was pre-trained on AudioSet and then fine-tuned on a small portion of the custom dataset (only ElevenLabs samples and real voices), demonstrated remarkable generalization capabilities. On the comprehensive test set, the model achieved an impressive EER of 0.91\%, indicating exceptional discriminative ability between bonafide and synthetic speech across all technologies.

\begin{table}[h]
\centering
\caption{Model Performance Across Voice Synthesis Technologies}
\label{tab:synthesis_performance}
\begin{tabular}{lc}
\toprule
Test Subset & Equal Error Rate (EER \%) \\
\midrule
Overall Test Set & 0.91 \\
ElevenLabs (seen during training) & 0.53 \\
NotebookLM (unseen) & 3.22 \\
Minimax AI (unseen) & 3.41 \\
Average on Unseen Technologies & 3.30 \\
\bottomrule
\end{tabular}
\end{table}

\begin{figure}[H]
  \centering
  \vspace{1em}
  \begin{minipage}{0.8\textwidth}
    \centering
    \includegraphics[width=\linewidth]{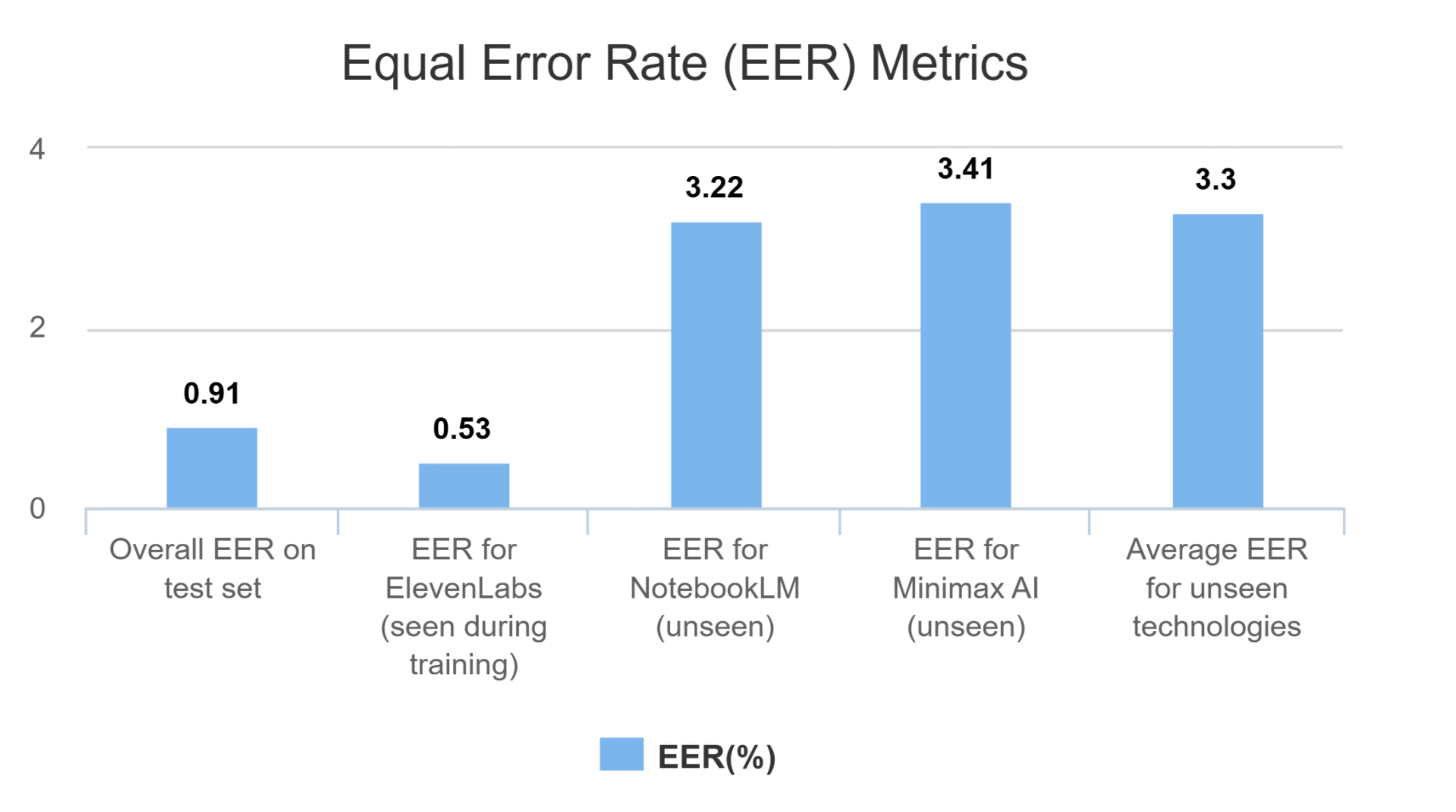}
    \label{fig:third}
  \end{minipage}
  \label{fig:all}
\end{figure}

These results highlight several critical insights:

\begin{enumerate}
    \item \textbf{Rapid Adaptation}: The AST architecture demonstrated an exceptional ability to quickly adapt to new voice synthesis technologies, requiring only 102 samples from a single synthetic voice generator (ElevenLabs) to achieve high performance.

    \item \textbf{Cross-Technology Transfer}: Training exclusively on ElevenLabs synthetic samples enables the model to detect other types of synthetic speech with reasonable accuracy (average EER of 3.3\% on unseen technologies), despite never being exposed to these technologies during training.

    \item \textbf{Architectural Advantages}: The transformer-based architecture appears particularly well-suited for identifying the subtle artifacts present in synthetic speech, regardless of the specific generation technology employed.
\end{enumerate}

The model's ability to generalize from a single synthetic voice technology (ElevenLabs) to other technologies (NotebookLM and Minimax AI) with an EER of 3.3\% suggests that certain common artifacts exist across different neural voice synthesis methods.

\section{Conclusion}
This research demonstrates the Audio Spectrogram Transformer's effectiveness in detecting synthesized speech across multiple generation technologies. The differentiated augmentation strategy for bonafide and synthetic speech successfully enhances the model's focus on persistent artifacts rather than dataset-specific characteristics. With minimal exposure to just one voice technology (102 ElevenLabs samples), AST achieves remarkable generalization, demonstrating 0.53\% EER on seen technologies and 3.3\% on unseen generators (NotebookLM and Minimax AI).

The results suggest common artifacts exist across different neural voice synthesis methods that transformer architectures can leverage for detection. This finding has significant implications as voice technologies evolve rapidly, indicating that effective countermeasures can be deployed even with limited training data from new generators.

Future work should expand evaluation to more voice synthesis technologies, investigate specific artifacts enabling cross-technology generalization, and develop more efficient fine-tuning approaches to further reduce adaptation samples required. This research establishes important benchmarks for rapid adaptation that can inform the development of more resilient speech verification systems in real-world applications where comprehensive training data may be unavailable.

Given the dynamic nature of deepfake audio generation, enterprises are advised to adopt a proactive defense strategy. This includes the utilization of anti-fraud services for real-time audio stream monitoring and the continuous refinement of deepfake detection databases to counter emerging synthetic audio technologies.

\section{Acknowledgments}
The work described in this paper was supported by ANTEI Limited AI Innovation Hub, a Hong Kong-based institution advancing the research and development of applicable AI technologies. The authors would also like to express their sincere thanks to the University of Information Technologies, Mechanics and Optics (ITMO University) and the Hong Kong University of Science and Technology (HKUST).

\begingroup
\sloppy

\end{document}